\documentclass[12pt]{article}
\usepackage{graphicx}
\usepackage{epsfig}
\usepackage{amsmath}
\usepackage{amsfonts}
\usepackage{graphics,color}
\def\scri{{\cal I}}
\def\bra#1{\left\langle #1\right\vert}
\def\ket#1{\left\vert #1\right\rangle}

\begin{document}
\begin{titlepage}
\rightline{\small{\tt MCTP-05-80}}
\rightline{\small{\tt NSF-KITP-05-30}}
\vspace{2 cm}
\begin{center}
{\Large\bf Beyond the Horizon}\\
\vspace{.7 cm}
{\sc Martin B Einhorn}\footnote{{\sl Email:}
{\tt ~meinhorn@kitp.ucsb.edu}}$^{,a,b}$
{{\sc and Manavendra Mahato}\footnote{{\sl Email:}
{\tt ~mmahato@umich.edu}}$^{,b}$
\\
\vspace{1 cm}
$^a${\sl Kavli Institute for Theoretical Physics,}\\
{\sl University of California,}\\
{\sl Santa Barbara, CA  93106-4030}\\
\vspace{.7 cm}
$^b${\sl Michigan Center for Theoretical Physics,}\\ 
{\sl Randall Laboratory, The University of Michigan,}\\
{\sl Ann Arbor, MI  48109-1120}}\\
\date{\today}
\end{center}
\vspace{1 cm}
\begin{abstract}
Cosmic horizons arise in general relativity in the context of black
holes and in certain cosmologies.  Classically, regions beyond a horizon
are inaccessible to causal observers.  However, quantum mechanical
correlations may exist across horizons that may influence local
observations.  For the case of de~Sitter space, we show how a single
particle excitation behind the horizon changes the density matrix
governing local observables.  As compared to the vacuum state, we
calculate the change in the average energy and entropy per unit volume. 
This illustrates what may be a generic property allowing some features
of spacetime beyond a horizon to be inferred.
\end{abstract}
\end{titlepage}
\pagenumbering{arabic}
\begin{section}{Introduction}

The analogies between thermodynamic concepts and black hole parameters
have posed numerous puzzles for quantum field theory (QFT) in curved
spacetime.  In particular, one would like to understand whether there is
a statistical mechanical origin of the Beckenstein\mbox{-Hawking}
entropy associated with the surface gravity or horizon of a black hole
\cite{B-H}.  This issue is complicated further by the fact that the
apparent horizon is frame dependent, so that the associated entropy is
also observer dependent.  Usually, discussions implicitly make reference
to stationary observers at infinity, and this frame dependence is
ignored. Similarly, Hawking's black hole information
paradox\cite{Hawking} deals with the changes experienced by external
observers who can measure black hole radiation. The challenges posed
have led some to speculate that, ultimately, local QFT as we know, must
break down even at distances large compared to the Planck length
\cite{some}. Indeed, the holographic principle suggests that quantum
field theory in general has many redundant degrees of freedom.
Similar paradoxes beset special reference frames or cosmological
situations in which there exist horizons, for which different observers
may describe things quite differently \cite{Unruh}. Such an observer
dependence was recently discussed by comparing observers in Rindler and
Minkowski reference frames \cite{MMR}.

Several approaches to these problems have illuminated the issues,
while not resolving them completely.  For eternal black holes, where
there is a delicately arranged balance between incoming and outgoing
radiation, a microscopic explanation is available, at least in some
cases, in terms of Israel's thermofield formulation\cite{Israel}.
Israel's construction, which may seem a bit artificial for ordinary
systems in flat spacetime, becomes natural for systems with horizons.
This is conceptually quite satisfying, inasmuch as the global
structure of the QFT is that of a unique state, whereas the entropy
ascribed to an observer is associated with the inability to access
states behind the horizon.

The case of de~Sitter space in static coordinates can be put into one-
to-one correspondence with Israel's discussion of the Schwarzschild
black hole \cite{BMS}, \cite{Spradlin:2001pw}.  An observer in the
southern diamond is classically blind to events occurring in the
northern diamond.  In this brief note, we will show that quantum
correlations provide a window for this observer to peek into the
northern diamond. In light of this result, we will reflect on some of
the so-called failures of QFT in this context.

In the next section, we review the construction of the vacuum state
for a scalar field in de~Sitter background.  We then consider the
effects of a single particle excitation in the northern diamond on
the mean energy (Section~3) and entropy (Section~4) associated with
an observer in the southern diamond.  We summarize our results and
discuss some of their implications in Section~5.

\end{section}
\begin{section}{Preliminaries}
To be self-contained, we will review in this section the description
of a scalar field in de~Sitter background.  We shall mostly follow the
notation in ref.~\cite{Spradlin:2001pw}.  
Four dimensional de~Sitter space can be thought of an embedding in (4+1)
dimensional Minkowskian space by a hyperboloid 
\begin{equation}
z_0^2-z_1^2-z_2^2-z_3^2-z_4^2=-{\alpha}^2.
\end{equation}
$\alpha$ is the de~Sitter radius, which henceforth will be set to unity.
To define a vacuum state for a free massless scalar field, one takes a
mode expansion
\begin{equation}
\phi=\sum_n [a_{n}{\phi}_{n}{(x)}+
a^{\dagger}_{n}{\phi}^{\ast}_{n}{(x)}]
\end{equation}
in terms of creation and annihilation operators with canonical
commutation relations (CCR). The vacuum is then defined as a state
destroyed by all the annihilation operators.
 \begin{equation}
a_{n}\ket{\Omega}=0
\end{equation}
The two-point function is defined as 
\begin{equation}
G_{\Omega}(x,x^{\prime})=
\bra{\Omega}\phi (x) \phi (x^{\prime})\ket{\Omega}  
= \sum_{n}{\phi}_{n}{(x)}{\phi}^{\ast}_{n}(x)
\end{equation}
The Euclidean counterpart of de~Sitter space is a four-dimensional
sphere $S^4.$  There is a unique choice of wave-function $ {\phi}_n$
that yields a Green's function nonsingular on $S^4.$ Correspondingly,
the two-point function obtained by the analytic continuation from the
Euclidean sphere is associated with a unique state called the Euclidean
vacuum $\ket{E}$. But one can define different vacua with respect to
different coordinate systems which appears equally `natural' in them.
For example, consider static coordinates in de~Sitter space. Regarded as
an embedding in five-dimensional Minkowskian space, they are defined in
the southern diamond ($0<r<1$) as
 \begin{eqnarray}
z_0&=&(1-r^2)^{1/2}\sinh(t)\\
z_1&=&(1-r^2)^{1/2}\cosh(t)\nonumber\\
z_2&=&r\sin\theta\cos\phi\nonumber\\
z_3&=&r\cos\theta\cos\phi\nonumber\\
z_4&=&r\cos\theta\nonumber
\end{eqnarray}
The line element takes the form
\begin{equation}
ds^2=[1-r^2]dt^2-[1-r^2]^{-1}dr^2-
r^2(d{\theta}^2+{\sin}^2{\theta}d{\phi}^2),
\end{equation}
with the timelike Killing vector $\partial_t$ defining time-evolution in
the southern diamond. The Penrose diagram for de~Sitter space is shown
in Fig.~1 \cite{Spradlin:2001pw}. 
\begin{figure}[hbtp]
     \centering
     \includegraphics[width=1.7truein]{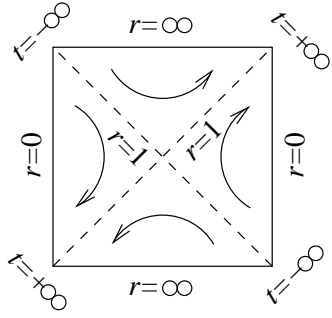}
     \caption[]{This Penrose diagram shows the direction of
     the flow generated by the Killing vector $\partial/\partial t$ in
     static coordinates.  The horizons (dotted lines) are at
     $r^2 =  1$, and the southern causal diamond
     is the region with $0 \le r \le 1$ on the right hand side.  Past
     and future null infinity $\scri^\pm$ are at $r = \infty$.}
     \label{static}
\end{figure}
These coordinates may be extended outside this causal diamond; for
$r>1,$ the roles of $t$ and $r$ are reversed, and $\partial_t$ becomes
spacelike. These coordinates describe only half of de~Sitter space, with
the remaining half described by replacing $t$ by $-t$ in the formulas
above. The Klein-Gordon equation in these coordinates is
\begin{equation}
\left \{-\frac{1}{1-r^2}{\partial}^{2}_{t}
+\frac{1}{r^2}{\partial}_{r}(1-r^2)r^2{\partial}_{r}
+\frac{1}{r^2 {\sin} \theta}
{\partial}_{\theta}(\sin\theta{\partial}_{\theta})
+\frac{1}{r^2 {\sin}^2 \theta }{\partial}^{2}_{\phi}
-m^2
\right \}\Phi=0
\end{equation}
It has a solution for a given energy $\omega$ and 
angular momentum quantum numbers $l,m$,  
\begin{eqnarray}
{\Phi}_{S,{\omega} ,lm}&=&f_{\omega ,lm}(r)
{\rm e}^{-i\omega t}Y_{l,m}(\theta,\phi)\\
f_{\omega ,lm}(r)&=&r^{l}(1-r^2)F(a,b,c,r^2)\nonumber\\
a&=&\frac{1}{2}(l +i\omega+h_{+}),\nonumber\\
b&=&\frac{1}{2}(l +i\omega+h_{-}),\nonumber\\
c&=&l+\frac{3}{2} .\nonumber
\end{eqnarray}
$Y_{l,m}(\theta,\phi)$ are the spherical harmonics and the function
$F(a,b,c,r^2)$ is a hypergeometric function. These form a
complete set of states for solutions regular at $r=0$. But they cannot
be extended to the whole of de~Sitter space. One can similarly define
modes in the northern diamond as
\begin{equation}
\label{phiN1}
{\Phi}_{N,{\omega} ,lm}= f_{\omega ,lm}(r){\rm e}^{-i\omega 
t}Y_{l.m}(\theta,\phi).
\end{equation}
One can consider the mode expansions
\begin{align}
\label{phiS}
{\Phi}_{S}&=\;\;{\int}^{\infty}_{0}d \omega 
\sum_{l,m}
a_{S, \omega ,lm}{\Phi}_{S, \omega ,lm}+
a_{S, \omega ,lm}^{\dagger}{\Phi}_{S, \omega ,lm}^*&\; {\rm in\;\; S}\\
&=0&\;{\rm in\;\; N}\nonumber\\
\label{phiN}
{\Phi}_{N}&=\;\;{\int}^{\infty}_{0}d \omega 
\sum_{l,m}
a_{N,\omega,lm}{\Phi}_{N, \omega ,lm} +
a_{N, \omega,l,m}^{\dagger}{\Phi}_{N, \omega ,lm}^*&\; {\rm in\;\; N}
\\      
&=0&\;{\rm in\;\; S}\nonumber.
\end{align}
On any spacelike slice through the origin, the $\Phi_S$ and $\Phi_N$
together form a complete set of functions for all of de~Sitter space.
Therefore, we may take $\Phi(x)=\Phi_S(x) + \Phi_N(x).$ The creation and
annihilation operators obey canonical commutation relations. Since modes
in one diamond have no support in the other, one can take the
corresponding operators to commute,
\begin{equation}
[a^{\dagger}_{N,\omega ,lm},a^{\dagger}_{S, \nu
,{l}^{\prime}{m}^{\prime}} 
]=0.
\end{equation}
As indicated in Fig.~1, the coordinate $t$ representing time runs in
forward direction in the southern diamond and backwards in the northern
one. So the modes in eq.~(\ref{phiN1}) are of negative frequency, so
that we must take $a_{N, \omega,lm}^{\dagger}$ as the annihilation
operator rather than $a_{N, \omega,lm}.$  The full Hamiltonian is
therefore difference $H=H_S-H_N$ (or vice versa.) 
One may then define a vacuum state as 
\begin{equation}
a_{S, \omega ,lm}\ket{ 0_{N,\omega  ,l,m}\:,\:0_{S,\omega  ,l,m}} =
a_{N, \omega,lm}^{\dagger}\ket{ 0_{N,\omega  ,l,m}\:,\:0_{S,\omega 
,l,m}}
=0.
\end{equation}
Henceforth, we shall simply abbreviate this vacuum,
which differs from the Euclidean vacuum, as $\ket{0}$ when
reference to the quantum numbers is not required. Note that the
associated Fock space is the direct product of the southern and northern
Fock spaces, ${\cal H}={\cal H}_S \otimes {\cal H}_N.$

\end{section}
\begin{section}{Energy change by particle excitation}

Components of ${\Phi}_{N}$ and ${\Phi}_{S}$ in equations ~(\ref{phiS})
and (\ref{phiN}) together account for a complete set of modes in
de~Sitter space. The Euclidean vacuum, when expressed in terms of these
states is
 \begin{equation}
\ket{E}= \prod_{\omega ,l,m}
(1-{\rm e}^{-2\pi\omega})^{1/2}
\exp[{\rm e}^{-\pi\omega}(a_{N,\omega  ,lm})(a_{S,\omega  
,lm}^{\dagger})]
\ket{0}
\end{equation}
Or, if we define $\ket{m_{N,\omega  ,l,m}\:,\:n_{S,\omega , l,m}}$ to
represent $m$ excitations of $(\omega, l,\;m)$ kind of modes in
northern diamond and $n $ excitations in southern diamond.
\begin{equation}
\ket{ m_{N,\omega  ,l,m}\:,\: n_{S,\omega  ,l,m}}=
\frac{(a_{N,\omega  ,l,m})^m}{\sqrt{m!}}
\frac{(a_{S,\omega  ,l,m}^{\dagger})^n}{\sqrt{n!}}
\ket{ 0 }
\end{equation} 
Then it can be shown that
\begin{equation}
\ket{E} = \prod_{\omega ,l,m}
(1-{\rm e}^{-2\pi\omega})^{1/2}
\sum_{n_{\omega  ,l,m}}^{\infty}
{\rm e}^{-\pi\omega{n_{\omega  ,l,m}}}\ket{ n_{N,\omega  
,l,m}\:,\:n_{S,\omega  
,l,m}}
\end{equation}
Measurements are classical, so an observer in the southern diamond
cannot directly probe states of the northern diamond. From the point of
view of a ``southern" observer, all observables can be obtained from the
density matrix formed by tracing over all the states corresponding to
the northern diamond modes.\cite{feynman}   Thus, a particular observer
actually appears to see a mixed state, even though the vacuum is a
unique state globally.
\begin{eqnarray*}
\rho^0_S&=&{\rm Tr}_{\rm N} \ket{E}\bra{E}\\
&=& \prod_{\omega ,l,m}
\sum_{k=0}^{\infty}
\bra{ k_{N,\omega ,l,m}}(\ket{E}\bra{E})
\ket{k_{N,\omega ,l,m}}
\end{eqnarray*}
Calculating the contribution for a given mode 
({\rm say,}\;$\omega=\nu,\;\{l,m\}=\{{l}^{\prime}{m}^{\prime}\}\;$),
\begin{eqnarray*}
\bra{k_N}
(1-{\rm e}^{-2\pi\nu})
\sum_{m=0}^{\infty}\;\;
\sum_{n=0}^{\infty}
{\rm e}^{-\pi\nu (m+n)}
\ket{m_N\:,\:m_S} \bra{ n_N\:,\:n_S}\ket{k_N} \\
=(1-{\rm e}^{-2\pi\nu})
{\rm e}^{-2\pi\nu k}
\ket{k_S}\bra{ k_S}
\end{eqnarray*}
This gives
\begin{equation}
\rho^0_S=(\rho^0_S)^{\prime}
\sum_{k=0}^{\infty}
(1-{\rm e}^{-2\pi\nu})
{\rm e}^{-2\pi\nu k}\ket{k_S}
\bra{ k_S}
\end{equation}
where
\[
(\rho^0_S)^{\prime}=
\prod_{\omega ,l,m}^{\prime}
\sum_{k_{\omega ,l,m}=0}^{\infty}
(1-{\rm e}^{-2\pi \omega })
{\rm e}^{-2\pi\omega k_{\omega ,l,m}}\ket{k_{S,\omega ,l,m}}
\bra{k_{S,\omega ,l,m}}
\]
Here, the prime over the product denotes that it is evaluated over all 
modes \underline{except} the particular mode ($\omega =\nu, 
\{l,m\}=\{{l}^{\prime},{m}^{\prime}\}$). 
One can check that the density matrix is properly normalised, {\it i.e.,}
${\rm Tr} \rho^0_S=1$.
Now suppose a single excitation of
$(\nu,\;{l}^{\prime},{m}^{\prime})$ 
mode of northern type is created in Euclidean vacuum. Since its 
wavefunction vanishes in southern diamond, one might naively expect 
the southern diamond observer to remain blind to this excitation. The
corresponding density matrix would be 
\begin{eqnarray*} 
\rho^1_S&=&{\cal N}{\rm Tr}_{\rm N}
(a_N\ket{E}
\bra{ E } a_N^{\dagger} ) \\
&=&{\cal N}(\rho^0_S)^\prime
(1-{\rm e}^{-2\pi\nu})
 \sum_{m=0}^{\infty}
 {\rm e}^{-2\pi\nu m}(m+1)
\ket{m_S}\bra{m_S}.
\end{eqnarray*}
To keep the density matrix normalised properly, the normalization factor
must be taken to be ${\cal N}=(1-{\bf  e}^{-2\pi\nu}).$  Therefore
\begin{equation}
\rho^{1}_S=(\rho^0_S)^{\prime}
(1-{\rm e}^{-2\pi\nu})^{2}
\sum_{m=0}^{\infty}
{\rm e}^{-2\pi\nu m}(m+1)\ket{ m_S}
\bra{  m_S} .
\end{equation}
The change in average energy in the southern diamond observer due to the
excitation of a northern diamond mode in Euclidean vacuum is therefore
\begin{eqnarray}
\Delta E_S&=&\nu {\rm Tr}[a_S^{\dagger}
a_S(\rho_S^{1}-\rho_S^0)]\nonumber\\
&=&\frac{\nu}{({\rm e}^{2\pi \nu }-1)}
\end{eqnarray}
which is precisely the energy $\nu$ of a single particle in a mixed
state corresponding to a Bose-Einstein distribution at the temperature
$(T=1/{2\pi})$.
\par
The energy change corresponding to an observer in the northern diamond
can be calculated in a similar fashion. The corresponding density matrix
for the case without any particles excited is  obtained by tracing over
all the states corresponding to the southern diamond.  
\begin{equation}
\rho^0_N={\rm Tr}_{\rm S}( \ket{E}\bra{E})
\end{equation}
The normalised northern diamond density matrix for one particle being 
created  would be
\begin{equation}
\rho^{1}_N=(\rho^0_N)^{\prime}
(1-{\rm e}^{-2\pi\nu})^{2}
\sum_{m=0}^{\infty}
{\rm e}^{-2\pi\nu (m-1)}m\ket{ m_N}
\bra{  m_N} 
\end{equation}
The change in energy corresponding to the northern diamond observer due
to the excitation of a northern diamond mode in Euclidean vacuum is
\begin{eqnarray}
\Delta E_N&=&\nu {\rm Tr}[a_{N,\nu l}
a_{N,\nu l}^{\dagger}(\rho_N^{1}-\rho_N^0)]\nonumber\\
&=&\frac{\nu}{(1-{\rm e}^{-2\pi \nu })}
\end{eqnarray}
One can check that $\Delta E_N-\Delta E_S=\nu $, the total change in
energy in de~Sitter space due to a single particle excitation in the
northern diamond.  

The lesson is that, because the Euclidean vacuum involves states in
which the southern and northern excitations are correlated, an
excitation in the northern diamond does have observable consequences in
the southern diamond, even though a southern observer cannot directly
probe northern states.  Of course, to interpret the density matrix,
a southern observer would need to know much about the global state, but
symmetry principles and the properties of quantum field theory go a long
way toward determining them.\cite{einhorn-larsen}

\end{section}
\begin{section} {Change in Entropy}
One can also calculate the entropy change observed by static observers
due to a northern diamond mode excitation. The entropy is simply $S_S=-
{\rm Tr}\rho_S\ln({\rho_S}).$  To evaluate this, it is convenient to
introduce some additional notation. Let $x \equiv {\rm exp}(-2\pi\nu),$
and let $A_S$ and $B_S$ denote the matrices corresponding to
$(\nu,l',m')$ submatrices of $\rho^0_S$ and  $\rho^{1}_S$ in the
southern diamond. Then
\begin{eqnarray}
A_S&=&(1-x) \sum_{n=0}^{\infty}
x^n \ket{n_S}\bra{n_S} \\
B_S&=&(1-x)^2\sum_{n=0}^{\infty}
x^n (n+1)\ket{n_S} \bra{n_S}
\end{eqnarray}
The change in entropy noted by southern diamond observer turns out to be
\begin{eqnarray}
\Delta S_S&=& -{\rm Tr}(B_S\ln{B_S}-A_S\ln{A_S})\nonumber\\
&=&-\ln(1-x)-\frac{x}{1-x}\ln{x} -
(1-x)^2\sum_{n=1}^{\infty}(n\ln{n}) x^{n-1}
\end{eqnarray}
The change in the entropy from the point of view of a northern diamond
observer is the same, since globally the system is in a pure state. This
is a special case of an important general theorem about the entropy of
factorized subsystems.  When a system is in a pure state, each subsystem
has the same set of nonzero eigenvalues, even if their dimension is
different.\footnote{See for example, \cite{Casini:2003ix}.}  Thus, even
though the excitations appear very different in the two causal diamonds,
and the change in energy in each is very different, the change in entropy 
is the same.

\end{section}
\begin{section}{Conclusion}

The Euclidean vacuum is a pure state with respect to the global
de~Sitter space.   Any observation can at best determine correlation
functions within a causal domain. The presence of cosmological horizons
in spacetime renders it impossible to measure correlations everywhere,
so all observables can be obtained from a density matrix formed by
tracing over states outside of the causal domain. Thus, any observer
perceives a mixed state.  In the case of the Euclidean vacuum, an
observer in static coordinates perceives a density matrix corresponding
to a thermal state.  

Naively, one might think that an observer in the southern diamond, for
example, would be insensitive to an excitation in the northern diamond
because its past support and influence in the future lie entirely outside the
southern diamond. However, because of correlations in the Euclidean
vacuum, such excitations change the density matrix for the southern
diamond and, therefore, can influence measurements there.  We have
illustrated this for the case of a single particle excitation in the
northern diamond, but the lesson is quite general.

These seemingly acausal results are reminiscent of other nonintuitive
correlations in quantum mechanics, as with the EPR paradox \cite{EPR}.
While we have discussed only the case of a particular cosmological
horizon, the same reasoning may help explain some of the properties of
black holes. Certainly, ``eternal" black holes, such as the classical
Schwarzschild geometry, will have their immediate counterparts to our
discussion.  It is perhaps not so incredible that a similar situation is
obtained for the case of gravitational collapse from a pure state to a
black hole followed by its subsequent evaporation due to Hawking
radiation.  These correlations across classical horizons may help
resolve the so-called black hole information paradox
\cite{Hawking}.
\end{section}
\section{Acknowledgments}
For permission to reproduce Fig.~1, we would like to thank the authors
of ref.~\cite{Spradlin:2001pw}. One of us (MBE) would like to
acknowledge helpful discussions with D. Marolf, A. Yarom, and R.
Brustein.  Much of this work was carried out while he was a participant
in the program {\it Superstring Cosmology}\/ at the KITP in the fall,
2003.  He would like to thank the coordinators for their hospitality
during his visit.  This research was supported in part by the National
Science Foundation under Grant No.~PHY99-07949 and by Department of
Energy Grant No.~DE-FG02-95ER40899.


\nocite{*}
\end{document}